\documentclass[showpacs,pra,twocolumn,floats,floatfix]{revtex4}
\usepackage{graphicx,epsfig}
\usepackage{times}
\usepackage{bm}
\usepackage{amsmath,amssymb}

\DeclareMathOperator{\re}{Re}
\DeclareMathOperator{\im}{Im}
\DeclareMathOperator{\var}{var}
\DeclareMathOperator{\tr}{tr}
\newlength\figurewidth
\begin{document}

\title{Opto-mechanical probes of resonances in amplifying
microresonators}
  \author{Henning Schomerus}
  \affiliation{Max-Planck-Institut f{\"u}r Physik komplexer Systeme,
  N{\"o}thnitzer Str. 38, 01187 Dresden, Germany}
  \author{Jan Wiersig}
  \affiliation{Institut f{\"u}r Theoretische Physik, Universit{\"a}t Bremen,
  Postfach 330 440, 28334 Bremen, Germany}
 \author{Martina Hentschel}
 \affiliation{Department of Physics, Duke University, Box 90305, Durham, NC 27708-0305}
\date{November 2003}

\begin{abstract}
We investigate whether the force and torque exerted by light pressure on an irregularly
shaped dielectric
resonator allow to detect resonant frequencies,
delivering information complemental to the scattering cross section
by mechanical means. The peak-to-valley ratio in the torque signal
can be many times larger than in the scattering cross section, and,
furthermore, depends on the structure of the resonance wave pattern.
The far-field emission pattern of the associated quasi-bound states
can be tested
by the angular dependence of the mechanical probes at finite amplification rate.
We relate the force and torque to the scattering matrix and
present numerical results for an annularly shaped dielectric resonator.
\end{abstract}
\pacs{03.65.Nk, 05.45.Mt, 42.25.-p, 42.60.Da}
\maketitle

\section{Introduction}

Waves confined in irregularly shaped geometries of optical microresonators
(such as micro-optical
lasers made of semiconductors~\cite{YS93},
micro-crystals~\cite{VKLISLA98},
or laser dye droplets~\cite{QSTC86})
pose practical and theoretical challenges, as
an intricate interference pattern arises from the
multiple coherent scattering off the confining boundaries.
This is especially true
at resonant conditions, when the multiple scattering results in
systematic constructive interference.
The most direct probes of these microresonators
are scattering experiments:
The systems are illuminated with a coherent light source,
and the scattering cross section is detected.
The resonant peaks observed in the cross section are related to
quasi-bound states (found at complex energies or frequencies),
which
can be observed as the working modes
of micro-optical lasers. Irregular geometries are
favored because they offer a rich mode structure and permit highly
anisotropic modes with well-defined
directed emission~\cite{GCNNSFSC98}.
The resonance pattern in the total scattering cross section 
does not reflect this
richness in the mode structure --- according to Breit-Wigner theory,
the weight and width of a resonance is determined
by the life-time of the mode; the wave pattern itself is of no concern \cite{petermann}.
In this paper, we show that complemental information about the
quasibound states is
contained in the  
directly accessible
opto-mechanical response of the system, which
actually {\it does} distinguish between modes of different
degree of anisotropy.

The opto-mechanical probes of the resonances and the associated
quasi-bound states that we discuss in this paper
are the force and the torque exerted on the
dielectric microresonator by the light pressure of an illuminating
beam.  We focus on the
practically most useful case of
irregular but effectively two-dimensional geometries,
and also allow for a finite amplification rate.
Over the past decade,
opto-mechanical tools based on light pressure
have found various applications
in the manipulation of microscopic objects,
for which usually simple shapes have been assumed,
and precision-detection of the acting forces have become commonplace \cite{ashkin}.
For instance, 
dielectric objects of micrometer dimensions
have been brought into rotation by light pressure, and the acting torque
has been determined from the rotation
of the object in a viscous medium
\cite{allen,yamamoto,friese,higurashi,galajda,santamato,luo}.
Typical torques are of the order of $10^{-17}$Nm for micrometer-sized objects
at 500nm
wave length and 10mW intensity.
These dimensions and operation parameters are also typical for optical
microresonators and microlasers.
The rotation rate
depends on the viscosity of the
ambient medium and is of order of several Hertz
(typical forces are of the order of $10-100$pN).
In some
of these experiments torques  as small as  $\simeq 10^{-19}$Nm proved to
be sufficient to induce rotation.
The rotation technique can be used not only to determine the torque,
but also to determine the viscosity once the torque has been obtained by independent
(e.g., optical) means \cite{nieminen,bishop1,bishop2}. 
Micro-electromechanical systems \cite{moreland} have not been applied in this
context so far, but in principle are also sensitive enough to detect the
opto-mechanical response.

\begin{figure}[t]
\includegraphics[width=0.4\figurewidth]{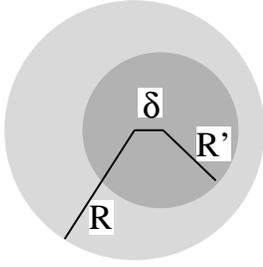}
\caption{The annular resonator used in the numerical investigations of this paper
is composed of two circles with radii $R$ and $R'=0.6\,R$, and
eccentricity $\delta=0.22\,R$.
The refractive index is $n=1$ outside the resonator, $n=1.8$ in the annular region
between the circles,
and $n=3.3$ inside the interior circle.
}
\label{fig:annbill}
\end{figure}

We suggest to use the opto-mechanical response for analyzing the
internal wave dynamics in complicated geometries.
We demonstrate
that the mechanical response contains information which is analogous to
the scattering cross section and the delay time
(the  conventional probes for resonances),
but is sensitive to other, complemental aspects of the quasi-bound states,
such as their degree of anisotropy (which is reflected by the torque).
Consequentially, the mechanical probes help to distinguish between
different wave patterns.
At finite amplification rate within the medium,
they contain information on the far-field emission pattern.
Our general argumentation is supported  in a practical setting by
numerical computations for the annularly shaped dielectric disk
shown in Fig.\ \ref{fig:annbill}, which
displays a multifaceted set of wave patterns due to
its non-integrable classical ray dynamics~\cite{BBEM93,HR02}.
We use two different numerical procedures, the wave-matching method
(see e.g. Ref.~\cite{HR02}) and the boundary element method~\cite{Wiersig02b}.

The organization of this paper is as follows: In Section \ref{secII},
we briefly discuss resonances, quasi-bound states,
and conventional probes of their
detection  (the Wigner delay time and the scattering cross section) in the
framework of scattering theory.
Section \ref{secIII} contains the principal results of this paper.
Section \ref{secIIIa} provides the kinematical
relations for the force and torque in
terms of the information provided by the scattering matrix.
In Section \ref{secIIIb} we  compare the
conventional and mechanical probes for the annular resonator
and show that the
anisotropy of the wave pattern is tested by the torque.
Rounding up the considerations, in Section \ref{secIIIc} we
discuss how the far-field emission pattern can be inferred
by means of amplification.
Our conclusions are collected in Section \ref{secIV}. The Appendix contains
details on the derivation of the kinematical relations.

\section{Resonances and quasi-bound states in scattering theory}
\label{secII}
\subsection{Scattering approach}
In this paper we apply standard scattering theory to
the effectively two-dimensional systems in question.
We separate the two polarizations
of the electromagnetic field, with either the electric
field
\begin{equation}
E_z=\re [\exp(-i\omega t)\psi]
\end{equation}
or the magnetic field
\begin{equation}
B_z=c^{-1}\re [\exp(-i\omega t)\psi]
\end{equation}
polarized perpendicular to the plane.
In both cases, the complex
wave function $\psi$ fulfills the
two-dimensional Helmholtz equation
\begin{equation}
(\Delta+n^2 k^2) \psi=0,
\label{helmholtz}
\end{equation}
where $n$
is the position-dependent refractive index, $k=\omega/c$
is the wave number, and $c$ is the speed of light in vacuum.
Amplification is modelled by a complex refractive index with $\im n<0$.

We choose a circular region $\cal A$ of radius ${\cal R}>R$, containing the resonator,
and decompose the wave function in the exterior of $\cal A$ in the
usual Hankel function basis,
\begin{equation}
\psi=\sum_{m} \left(a^{(\rm in)}_m
H^{(2)}_m (k r)+ a^{(\rm out)}_m
H^{(1)}_m (k r)\right)e^{i m\phi}
,
\label{eq:hankel}
\end{equation}
where $r$ and $\phi$ are polar coordinates.
(Here and in the following, all sums run from $-\infty$ to $\infty$.)
The angularly resolved radiation in the far field is given by
\begin{equation}
I(\phi)=\frac{2}{\pi |k|}\left|\sum_{m} a^{(\rm out)}_m e^{i m(\phi-\pi/2)}\right|^2
.
\label{eq:I}
\end{equation}

Below the laser threshold,
the expansion coefficients  $a^{(\rm out)}_m$ of the
outgoing wave are related to their incoming
counterparts $a^{(\rm in)}_m$ by linear relations
\begin{equation}
a^{(\rm out)}_m = \sum_{m'} S_{mm'}a^{(\rm in)}_{m'},
\label{eq:srel}
\end{equation}
where the coefficients $S_{mm'}$ form the scattering matrix.
The scattering matrix fulfills
the time-reversal symmetry
\begin{equation}
S_{m,n}=S_{-n,-m}(-1)^{m+n},
\end{equation}
and is unitary for real $k$ and $n$. For complex values, unitarity is
replaced by
\begin{equation}
S^{-1}(k,n)=[S(k^*,n^*)]^\dagger.
\end{equation}

\subsection{Poles and quasi-bound states}

Quasi-bound states are found at complex values $k_c$ of $k$ that permit
nontrivial solutions ${\bm a}^{(\rm out)}={\bm a}_c$
in the case of no incident radiation,
\begin{equation}
{\bm a}^{(\rm in)}={\bm S}^{-1} {\bm a}^{(\rm out)}=0.
\end{equation}
It follows that the values $k_c$ are the poles of the scattering matrix,
which (for real $n$) all reside in the lower complex plane as a consequence of causality.

For the annular resonator, complex values $k_c$ are shown in the bottom panel of Fig.\ \ref{fig:s1},
and some typical wave patterns are presented in the right panels of Fig.\ \ref{fig:angular}.
For this system with a reflection symmetry about the $x$ axis,
the quasi-bound states occur
with even and odd parity \cite{frischat}, and further
can be divided into whispering-gallery modes localized at the interior interface (class $\rm W_{int}$)
or at the exterior interface (class $\rm W_{ext}$), and more extended modes
that we group into two classes $\rm A$ (inside the interior circle, these
modes are of whispering-gallery character)
and $\rm C$ (the remaining modes).
Close to resonance in the complex $k$-plane ($k\approx k_c^{(\rm even)}, k_c^{(\rm odd)}$), the scattering matrix can
be approximated by \cite{petermann2}
\begin{equation}
S\approx \frac{{\bm a}^{(\rm even)}_c\otimes
\tilde{\bm a}^{(\rm even)}_{c}}{k-k_c^{(\rm even)}}+
\frac{{\bm a}^{(\rm odd)}_c\otimes
\tilde{\bm a}^{(\rm odd)}_{c}}{k-k_c^{(\rm odd)}}
\label{eq:sres}
\end{equation}
where $\tilde a_{c,m}=(-1)^m a_{c,m}$ due to time-reversal symmetry
and we accounted for the potentially
quasi-degenerate partner of opposite parity (indicated by the superscripts).

In general, the quasi-bound states found for complex $k_c$ and real $n$ can be
transported to real values of $k\simeq\re k_c$ 
by setting $\im n\simeq\im k_c\re n/\re k_c$, corresponding
to an active, amplifying medium close to threshold \cite{remark0}.
Above threshold, poles formally move into the upper complex plane, which physically indicates
instability, and the
linear relation between  ${\bm a}^{(\rm in)}$ and  ${\bm a}^{(\rm out)}$
breaks down.
Yet,  in homogeneously
amplifying media the lasing modes are well
approximated by the cold-resonator modes \cite{laserbooks,beenakker,cao}.
The far-field emission pattern of the laser,
\begin{equation}
I_c(\phi)=\frac{2}{\pi |k|}\left|\sum_{m} a^{(\rm out)}_{c,m} e^{i m(\phi-\pi/2)}\right|^2
,
\label{eq:ic}
\end{equation}
is then given by Eq.\ (\ref{eq:I}),
evaluated with the quasi-bound state ${\bm a}_c$ that wins the mode competition
(the first mode that becomes unstable).

\subsection{Resonances and their conventional probes}

The values $k_c$ are the poles of $S$
where this matrix is singular, and are reflected by
resonances of the system at real values $k=\re k_c$.
Conventional probes for resonances are obtained from the
total scattering cross section
\begin{equation}
\label{eq:sigma}
\sigma=\frac{4}{k}\sum_{m}|a_m^{(\rm out)}-a_m^{(\rm in)}|^2
,
\end{equation}
and the
weighted delay time \cite{delaytimes}
\begin{equation}
\label{eq:tau}
\tau=4 \im \sum_m a_m^{(\rm out)*}da_m^{(\rm out)}/d\omega
.
\end{equation}
Clearly, the quantities $\sigma$ and $\tau$
depend on the incoming wave, which we now
specify as a plane wave
coming from direction $\phi_0$ \cite{remark},
corresponding to
\begin{equation}
a^{(\rm in)}_m=\frac{1}{2}e^{-im(\pi/2+\phi_0)}.
\label{eq:ainphi}
\end{equation}
Both $\sigma$ and $\tau$ then provide angularly resolved information
as a function of $\phi_0$.
A global characterization of the system is obtained by an
average over the incident radiation direction
$\phi_0$, giving
\begin{eqnarray}
&&\sigma_0=\langle\sigma\rangle_{\phi_0} =
\frac{1}{k}\sum_m (1+\sum_{m'}|S_{m'm}|^2-2 \re S_{mm}),\quad\,\,\,
\\
&&\tau_{\rm W}=\langle\tau\rangle_{\phi_0} =\im \tr S^\dagger \frac{dS}{d\omega},
\end{eqnarray}
where $\tau_{\rm W}$ is known as the Wigner delay time.

\begin{figure}[t]
\includegraphics[width=0.9\figurewidth]{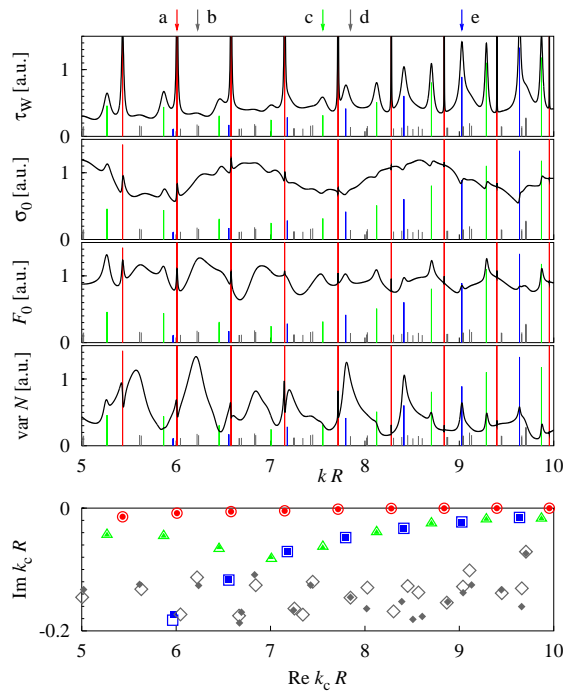}
\caption{(Color) The top panels show four quantities
(in arbitrary units) as a function
of $k$ that probe for resonances in the annular resonator: The
Wigner delay time $\tau_{\rm W}$,
the angle-of-incidence averaged
scattering cross section $\sigma_0$, the angle-of-incidence averaged
force in forward direction $F_0$,
and the variance of the torque $\var N$.
The lower panel shows the complex resonance wave numbers $k_c$ of quasi-bound states,
classified by their wave patterns as described in the text:
${\rm W}_{\rm int}$ ($\circ$),
${\rm W}_{\rm ext}$ ($\triangle$),
${\rm A}$ ($\Box$),
${\rm C}$ ($\diamond$).
Open symbols indicate modes of even parity, full symbols are modes of odd
parity.
The complex resonances
are also indicated by the spikes in the top
panels, located at $k=\re k_c$  with height
proportional to the life time $1/(-2c\im k_c)$.
Additionally, the resonances of Fig.\ \ref{fig:angular}(a-e)
are indicated by the arrows at the very top.
}
\label{fig:s1}
\end{figure}

Both the delay time (of predominantly theoretical virtue)
and the (more practical) scattering cross section
display peaks at resonance, as is illustrated for the annular resonator
in the two topmost graphs of Fig.\ \ref{fig:s1}.
Note that the signal
$\sigma_0$ is, in general, of low contrast and, apart from a
relatively strong
background modulation, rather featureless. Still,
the enhanced scattering of the light field at resonance
promises a marked mechanical response,
which we investigate in the remainder of this paper.
Indeed, the signal of the mechanical response will display a much 
better contrast for a set of systematically selected resonances.

\section{Mechanical detection of resonances}
\label{secIII}
\subsection{Kinematics}
\label{secIIIa}
In this subsection we provide general kinematic relations for
a two-dimensional resonator in a light field,
which in the following Subsections \ref{secIIIb},\ref{secIIIc} will be used to
characterize resonances and wave patterns.

The mechanical forces exerted by the light field on a dielectric medium originate from
the refraction and diffraction at the dielectric interfaces ---
ultimately, from the deflection (and creation, at finite amplification)
of the  photons.
The kinematics in the combined system of light field and
medium can be obtained
from the conservation  laws of total
angular and linear momentum,
which equate the torque and force acting on the medium to the
deficit of the angular and linear
momenta carried  by the electromagnetic field into and out of the
circular region $\cal A$.
(The center of this region is identified with the point of reference
for the torque, and in our example is taken as the center of the exterior circle.)
Note that the conservation laws also hold in an amplifying medium, due
to the recoil of each created photon.
The kinematic relations hence follow
from integrals of Maxwell's stress tensor~\cite{Jackson75} over the boundary of $\cal A$.
After some algebra (for details see the Appendix),
we find the time-averaged force and torque
(per unit of thickness of the resonator) in
 the compact form \cite{barton}
\begin{eqnarray}
F_x+i F_y&=&\frac{2 \varepsilon_0 i}{k}\sum_m
(a_m^{(\rm in)}a_{m+1}^{{(\rm in)}*}
- a_m^{(\rm out)}a_{m+1}^{{(\rm out)}*})
,
\label{eq:force}
\\
N&=&\frac{2\varepsilon_0}{k^2}
\sum_m m \left(|a^{(\rm in)}_m|^2-|a^{(\rm out)}_m|^2 \right)
.
\label{eq:torque}
\end{eqnarray}
In absence of amplification and for
plane-wave illumination,
it is easily seen that the direction-averaged force and
torque
vanish from the unitarity constraints of the scattering matrix:
\begin{eqnarray}
\langle F_x+i F_y\rangle_{\phi_0}
=
\frac{- i\varepsilon_0}{2k}\sum_{m,m'} S_{mm'}S^*_{m+1,m'}
&=&0,
\\
\langle N \rangle_{\phi_0}
=
\frac{\varepsilon_0}{2k^2}\sum_m m
\left( 1-\sum_{m'}|S_{mm'}|^2\right)&=&
0.\quad
\end{eqnarray}
Two simple quantities that do not vanish are the mean of the force component
$F_\parallel = F_x \cos\phi_0+ F_y\sin\phi_0$
in forward direction,
\begin{equation}
F_0 =\langle F_\parallel \rangle_{\phi_0} =
\frac{\varepsilon_0}{2 k}\re \sum_{m,m'}(\delta_{m m'}-S_{m'm}S^*_{m'+1,m+1})
,
\end{equation}
and the variance $\var N=\langle N^2\rangle_{\phi_0}$ of the torque,
\begin{equation}
\var N=\frac{\varepsilon_0^2}{4k^4}
\sum_{m,m',n_i}'mm'S_{m n_1}^*S_{m n_2}S_{m' n_3}^*S_{m'n_4}.
\end{equation}
Here, the prime at the sum enforces the restriction $n_1+n_3=n_2+n_4$.

\subsection{Resonances}
\label{secIIIb}

\begin{figure}
\includegraphics[width=0.916\figurewidth]{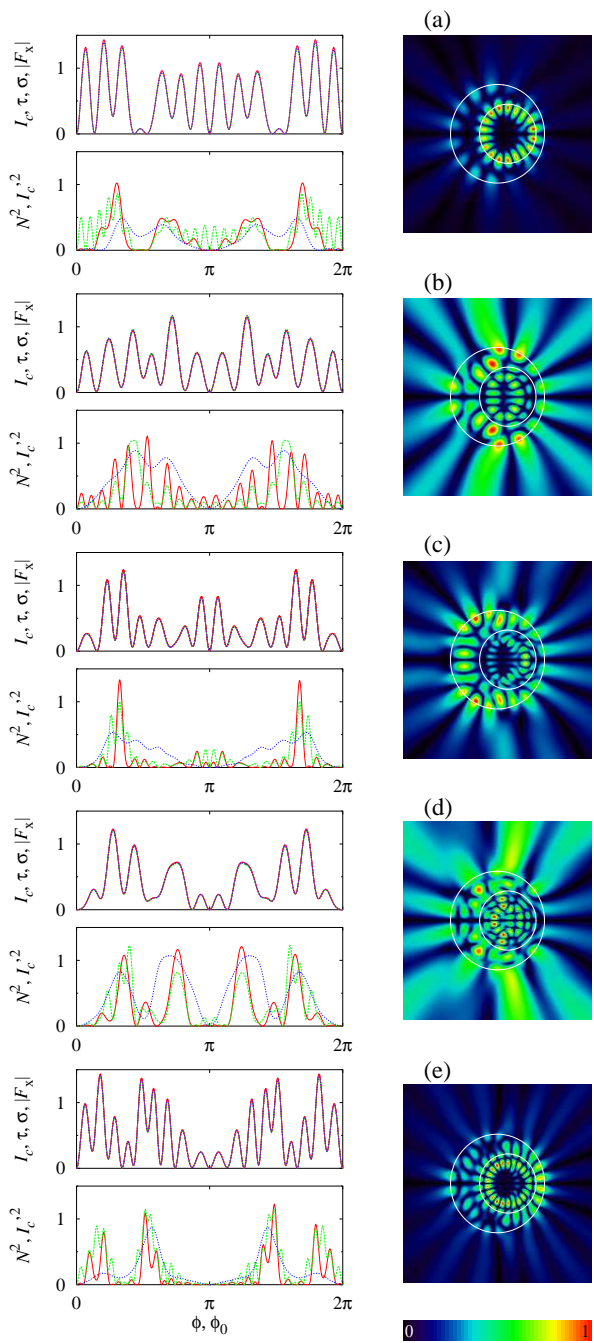}
\caption{(Color)
The panels (a-e) show results for various combinations of $k$ and $\im n$
tuned very close to resonance with the quasi-bound states of odd parity
shown on the right, with
(a) $k_c=6.009-i\,0.008$ (class $\rm W_{int}$),
(b) $k_c=6.228-i\,0.116$ (class $\rm C$),
(c) $k_c=7.554-i\,0.063$ (class $\rm W_{ext}$),
(d) $k_c=7.847-i\,0.145$ (class $\rm C$),
(e) $k_c=9.026-i\,0.022$ (class $\rm A$).
In the top graph of each  panel,
the angular dependence of the far field $I_c$ (red)
on the radiation direction $\phi$ is compared to the angular dependence on the
illumination direction $\phi_0$ (note Ref.\ \cite{remark})
 of the scattering cross section
$\sigma$ (green),
the weighted delay time $\tau$ (blue), and the
 $x$-component of the force,
$F_{x}$ (purple) --- the four lines
are almost indistinguishable.
In the bottom graph of each panel, the (squared) torque $N^2$
(red) is compared to the interference
term $I_c^{\prime 2}$ (green),
defined in Eq.\ (\ref{eq:ir}).
The blue curve is $N^2$ evaluated at real $n$.
All quantities are in arbitrary units.
}
\label{fig:angular}
\end{figure}

Because scattering is enhanced at resonance,
we expect that
$\var N$ and $F_0$ are global characteristics of the resonances
comparable to $\sigma_0$ and $\tau_W$, while $N$ and $F_\parallel$ provide
angularly resolved information
as a function of the incident radiation direction $\phi_0$,
analogously to $\sigma$ and $\tau$.
The results in Figs.\ \ref{fig:s1} and \ref{fig:angular} demonstrate
this promise to hold true.

Figure \ref{fig:s1} shows
the global characteristics
$\sigma_0$,  $\tau_W$, $\var N$, and $F_0$
for the annular resonator
as a function of $k$.
The plot demonstrates
a clear correspondence of the resonant peaks in all four quantities.
Evidently, each quantity probes another
aspect of the resonances, such that their relative weights are different.
As usual, the delay time $\tau_W$ displays the largest peaks for the very narrow resonances
associated to long-living quasi-bound states.
Presently, the longest-living states are those of class
$\rm W_{int}$, followed by
those of class $\rm W_{ext}$ and $\rm A$, while the states of class $\rm C$ are hardly visible here.
The scattering cross section $\sigma_0$
displays smaller peaks (sometimes, dips) on a modulated
background, and does not
provide a distinctive discrimination between the different states.

The force and torque are sensitive to the wave pattern itself,
and for the annular resonator display a marked response especially for the
quasi-bound states of class $\rm W_{ext} $ and $\rm A$
(for small $k$, also for class $\rm C$).
The peak-to-valley ratio in the variance of the
torque (e.g., $\approx 5$ for the peak at $k\approx 6.2$)
can take much larger values than in the scattering cross section
(where the ratio rarely exceeds 1.2) for resonances with a rather
anisotropic internal wave pattern.

This principal conclusion
of the present paper is supported in more detail by the
five examples in Fig.\ \ref{fig:angular}.
The wave patterns shown in the right panels
are inhomogeneous to various extent,
even though the inhomogeneity does not automatically
translate into very anisotropic far-field emission patterns
for these rather low values of $\re k_c R$.
Let us inspect the associated resonance peaks  in Fig.\ \ref{fig:s1}.
For the quasibound state of Fig.\ \ref{fig:angular}(a), force, torque, and
scattering cross section detect the resonance with comparable contrast;
the best signal is given by the delay time, which, however, must be remembered to be an
inconvenient tool for practical considerations.
Figure \ref{fig:angular}(b) pertains to a comparatively short-living
mode (with large $|\im k_c|$).
This mode is hard to detect by the conventional probes,
but gives a very clear signal in the mechanical probes,  especially, in
the torque.
For the case of Fig.\
\ref{fig:angular}(c), the best signals are provided
by the force and by the delay time,
followed by the scattering cross section, while the torque is
insensitive to this resonance. The case of Fig.\ \ref{fig:angular}(d)
illustrates that resonances with both a short life time
and a rather homogeneous wave pattern
are hardly detected by any of the four probes --- such modes, however,
are also of subordinate interest in the practical applications. 
The mode of  Fig.\ \ref{fig:angular}(e) gives a torque signal with a 
much higher contrast than the force or the scattering cross section;
only the delay time signals the resonance to a
similar extent.

The most remarkable difference in Fig.\ \ref{fig:angular} is between
panels (b) and (d)
--- both modes of class C displayed there have a short life time, which
inhibits their detection by conventional means,  
but the mode of panel (b) is detectable by opto-mechanical means,
in keeping with the larger anisotropy displayed by its wave pattern
(note the regions of concentrated intensity above and below the interior circle).

\subsection{Far field}
\label{secIIIc}

In the practical application of micro-optical lasers, the
anisotropy of the internal wave pattern is a desired  design goal
since it is the pre-requisite
for directed emission in the far field.
To which extent can the far field be inferred from the
angularly resolved information contained in $N$, $F$, $\sigma$, and $\tau$?
This rather innocent question turns out to be surprisingly subtle in view
of the following observation: In absence of amplification
(real $k$ and $n$),
the angle-of-incidence averaged far
field
\begin{equation}
I_0(\phi)=\langle I(\phi)\rangle_{\phi_0}
\label{eq:io}
\end{equation}
is independent of $\phi$.
Here, we average over the angle of incidence $\phi_0$ since the
quasi-bound states are defined without reference to any excitation mechanism.
The angle of incidence $\phi_0$ enters the far field
(\ref{eq:I}) via Eq.\ (\ref{eq:ainphi}), and
after taking the average the statement above follows
from the unitarity of the scattering matrix.
In other words, in the absence of amplification,
the angle-of-incidence averaged
far-field radiation pattern does not carry any intrinsic information
about the quasi-bound states, even at resonant conditions.

Can the far field of the quasi-bound states be inferred from a more sophisticated 
analysis of $I(\phi;\phi_0)$, i.e., by taking the dependence of the angle of incidence
into account? This would require the delicate task to discard
the component of the light that is directly reflected at the first encounter
of the boundary from the outside. The direct contribution is essentially independent
of the quasi-bound states: the latter are determined
by a constructive-interference condition for reflection from the inside
of the system while the directly reflected radiation never ever enters the medium.
For instance, in our model system, the directly reflected wave component
contains no information on the interior circle, which, however, 
is crucial in the formation of  all quasi-bound states.

The situation changes significantly
at {\em finite amplification}: Then, already the angle-of-incidence averaged intensity
$I_0(\phi)$ is modulated,
and
is influenced
by the quasi-bound states closest in $k$.
At exact resonance in the complex $k$-plane,
approximation (\ref{eq:sres})  of the scattering matrix
entails ${\bm a}^{(\rm out)} \propto {\bm a}_c$ with a large
proportionality constant due to the resonant denominator.
Hence the contribution ${\bm a}^{(\rm in)}$ of the incident radiation 
can be neglected, and
$I(\phi)\propto I_c(\phi)$ as defined in Eq.\ (\ref{eq:ic}),
independent of the mode of excitation.
Furthermore
\begin{equation}
\sigma(\phi_0)\propto\tau(\phi_0)\propto |F_x(\phi_0)|\propto I_c(\phi=\phi_0),
\label{eq:corres}
\end{equation}
since again ${\bm a}^{(\rm out)}  \propto {\bm a}_c$ dominates over 
${\bm a}^{(\rm in)}$ in Eqs.\ (\ref{eq:sigma}),
(\ref{eq:tau}), (\ref{eq:force}).
Equation (\ref{eq:corres}) entails a
duality between the illumination direction $\phi_0$ and
the radiation direction $\phi$. This duality
relies on the time-reversal symmetry \cite{remark},
which is incorporated in Eq.\ (\ref{eq:sres})
by the relation between ${\bm a}_c$ and $\tilde{\bm a}_c$.
For a representative set of resonances at real $k$ and complex $n$,
the far field $I_c(\phi)$ 
is shown in the left panels of
Fig.\ \ref{fig:angular}, along with the angular dependence of
$\sigma$, $\tau$, $|F_x|$, and $N$ on the direction of incident radiation,
$\phi_0$. The wave pattern of the corresponding
quasi-bound states is shown in the right panels.

Due to the reflection symmetry of the annular resonator about the $x$ axis,
$|F_y|\ll |F_x|$ while $F_{||} \approx \cos(\phi_0)F_x$.
The reflection symmetry also suppresses $N$
{\em compared to non-symmetric systems}:
The otherwise dominant contribution from ${\bm a}^{(\rm out)}={\bm a}_c$  vanishes for
each given quasi-bound state,
$\sum_m m|a_{c,m}|^2=0$, because of pairwise cancellation
of the terms with opposite $m$.
However,
$N$ is still enhanced {\em compared to the non-resonant situation}:
A non-vanishing result (of order $|k-k_c|^{-1}$)
is obtained from the interference between the resonant state
(with large coefficients) and the non-resonant states
(with moderate coefficients).
In the typical case of quasi-degeneracy,
${\bm a}^{\rm (even)}_c$ interferes
with ${\bm a}^{\rm (odd)}_c$, and from Eq.\ (\ref{eq:sres}) we obtain
\begin{equation}
N^2(\phi_0)\propto I_c^{\rm (even)}(\phi_0)I_c^{\rm (odd)}(\phi_0) \equiv
I_c^{\prime 2}(\phi_0)
.
\label{eq:ir}
\end{equation}
This relation is indeed obeyed to good extent in the numerical computations (see Fig.\ \ref{fig:angular}).
Even at real $n$, $N^2$ (the blue curve in the bottom graph of each panel) roughly corresponds to $I_c^{\prime 2}$.
In non-symmetric geometries the sum $\sum_m m|a_{c,m}|^2\neq 0$
for a given quasi-bound state, and hence
the proportionality $N\sim I_c$ (and $F_y\sim I_c$)
is restored (moreover, quasi-degeneracies are then lifted, and the system
more easily is tuned to resonance with individual quasi-bound states).

\section{Conclusions}
\label{secIV}

In summary, the force and torque
exerted by light pressure on a dielectric resonator allow
to detect resonances and help to characterize the wave patterns
of the associated quasi-bound states.
For anisotropic wave patterns that support a high angular-momentum transfer,
the peak-to-valley ratio in the mechanical probes (notably the torque)
exceeds by far the moderate values observed in the scattering cross section,
which is notoriously insensitive to the wave pattern.

We put our work in the context of directed transmission from micro-optical lasers,
and took amplification into account. Enhanced sensitivity to internal
structure is also desired in several applications involving simpler passive or
absorbing media, such as cells with organelles or liquid drops polluted with inclusions.

Since resonance provides a very effective scattering mechanism
by constructive interference,
the typical forces and torques estimated for common micrometer sized
dielectric resonators under typical radiation conditions are in the range of
the opto-mechanical experiments mentioned in the introduction.
In the viscous rotation experiments, the resonances
will depend on the refractive index of the surrounding liquid, which may not
be desirable for a precise characterization of the resonator,
but also introduces an additional
potentially useful control parameter. Detection by microelectromechanical systems
offers the advantage to
control the resonator orientation with respect to the incoming
radiation.

The numerical part of this work concentrated on the wave-optical regime,
in which the wave length is not much smaller than the
geometric features of the system and interference patterns are most
complex.
Some micro-optical lasers operate at smaller wave lengths
(larger values of the wave number $k$) than accessed in our numerics.
In this regime, semiclassical relations
between the
internal wave pattern and the far-field emission pattern can be
formulated, which also relate the
directed emission desired for micro-optical laser to the underlying
anisotropy of the internal wave pattern.
The results of this work hold the promise that in this semiclassical regime,
modes with directed emission can be identified by the torque.
The selectivity of the torque for anisotropic modes should be even
enhanced for these larger values of $k$ by the following mechanism:
Isotropic modes frequently
arise from a large collection of unstable ray trajectories \cite{Gutz90},
and can be described as a superposition of random waves \cite{Berry77}.
As $k$ is increased, more and more random-wave components become available,
and hence the mechanical response is suppressed by self-averaging.
Anisotropic modes are guided by just a few trajectories
such that the self-averaging mechanism does not apply to them,
and consequently they remain well detectable by their opto-mechanical response.

\appendix*
\section{Derivation of the kinematical relations}

The calculation of the force
\begin{equation}
{\bf F}=\int_S dS\, {\bf n} T
\end{equation}
and the torque
\begin{equation}
{\bf N}=\int_S dS\, ({\bf n} T)\times {\bf r}
\end{equation}
(where $\bf n$ is the unit vector in normal direction to the surface $S$)
starts with the stress tensor
\begin{equation}
T=\varepsilon_0({\bf E}\otimes{\bf E}-\frac{1}{2}E^2\openone)
+\frac{1}{\mu_0}({\bf B}\otimes{\bf B}-\frac{1}{2}B^2\openone)
.
\end{equation}
In the two-dimensional
case, we integrate over the circle $\partial A$ of radius ${\cal R}$
(the physical force and torque are obtained by a multiplication with the
thickness of the sample), and it is natural to work in polar coordinates.

Depending on the polarization, we insert the electromagnetic field
\begin{eqnarray}
&&\left.\begin{array}{l}
{\bf E}=\re e^{-i\omega t}\psi{\bf e}_z,
\\
{\bf B}=\im \omega^{-1}e^{-i\omega t}
({\cal R}^{-1}\partial_\phi\psi{\bf e}_r- \partial_r\psi{\bf e}_\phi),\end{array}\right\}\quad\mbox{(TM)},
\nonumber
\\
&&\left.
\begin{array}{l}
{\bf B}=c^{-1}\re e^{-i\omega t}\psi{\bf e}_z,
\\
{\bf E}=-c\im \omega^{-1}e^{-i\omega t}
({\cal R}^{-1}\partial_\phi\psi{\bf e}_r- \partial_r\psi{\bf e}_\phi),\end{array}\right\}\quad\mbox{(TE)},
\nonumber
\\
\end{eqnarray}
where the wave function fulfills the Helmholtz equation (\ref{helmholtz})
and is decomposed in the basis of Hankel function, Eq.\ (\ref{eq:hankel}).

After a time-average,
the radial component of the stress tensor then  is given in terms of
$\psi$ by
\begin{eqnarray}
 {\bf n} T&=&
\frac{\varepsilon_0}{4}
\left[ (Z^{-2}|\partial_\phi \psi|^2-|\psi|^2- |\partial_Z \psi|^2){\bf
e}_r
\right.
\nonumber \\
&&\left.
 -2Z^{-1}\re\partial_\phi\psi\partial_Z\psi^*{\bf e}_\phi
\right]
,
\end{eqnarray}
with $Z=k {\cal R}$, $\partial_Z = \partial_{kr}|_{r={\cal R}}$,  $k=\omega/c$, $c=1/\sqrt{\mu_0 \varepsilon_0}$.
(Note that $k$ is real and $n=1$ on $\partial A$,
since refraction and amplification is restricted to the resonator.)

The force components can now be expressed as
\begin{eqnarray}
F_x+i F_y&=&
\frac{\varepsilon_0 {\cal R}  }{4} \int_0^{2\pi} d\phi e^{i\phi}
\left( \frac{1}{Z^2}|\partial_\phi \psi|^2- |\partial_Z \psi|^2\right.
\nonumber \\
&&\left.
-|\psi|^2
 -\frac{i}{Z}\partial_\phi\psi\partial_Z\psi^* -\frac{i}{Z} \partial_\phi\psi^*\partial_Z\psi\right)
.
\qquad
\end{eqnarray}
Here we insert Eq.\ (\ref{eq:hankel})
[indices $m,j$ for $\psi$, indices $m',j'$ for $\psi^*$, where the
second index $j,j'=(1)\equiv \rm out$ or $(2)\equiv \rm in$]
and integrate over $\phi$.
Next, we use the identities
\begin{eqnarray}
\partial_Z H_m^j&=&\frac{m}{Z} H_m^j-H_{m+1}^j,
\\
\partial_Z H_{m+1}^{j'*}&=& -\frac{m+1}{Z}H_{m+1}^{j'*}+H_m^{j'*}
\end{eqnarray}
(here and in the following, we suppress the argument $Z$ of the Hankel
functions).
This gives
\begin{eqnarray}
F_x+i F_y&=&
\frac{\pi \varepsilon_0 {\cal R}  }{2}
\sum_{m,j,j'}
a_m^ja_{m+1}^{j'*}
\nonumber\\
&&
\left[
\Big(H_{m+1}^j-\frac{m}{Z} H_m^j\Big)\Big(
H_m^{j'*}-\frac{m+1}{Z}H_{m+1}^{j'*}
\Big)
\right.
\nonumber\\
&&
+\frac{m(m+1)}{Z^2}H_m^jH_{m+1}^{j'*}
-H_m^jH_{m+1}^{j'*}
\nonumber\\
&&
-\frac{m+1}{Z}\left(\frac{m}{Z} H_m^j-H_{m+1}^j\right)H_{m+1}^{j'*}
\nonumber\\
&&
\left.
+\frac{m}{Z}
H^j_m\left(-\frac{m+1}{Z}H_{m+1}^{j'*}+H_m^{j'*}\right)
\right]
.
\end{eqnarray}
Between the square brackets, most terms cancel, giving
\begin{eqnarray}
F_x+i F_y&=&
\frac{\pi \varepsilon_0 {\cal R}  }{2}
\sum_{m,j,j'}
a_m^ja_{m+1}^{j'*}A_{j,j'}
,
\nonumber\\
A_{j,j'}&=&\left( H_{m+1}^jH_{m}^{j'*}-H_m^jH_{m+1}^{j'*}\right)
.
\end{eqnarray}
With $H^{(1)*}=H^{(2)}$ and the
identities
 \begin{equation}
A_{11}=-A_{22}=2i\im H_{m+1}^{(1)}H_m^{(2)}=-\frac{4i}{\pi Z},
\end{equation}
$A_{12}=A_{21}=0$,
we arrive at the final expression (\ref{eq:force}).

The calculation for the torque is less involved.
We find
\begin{eqnarray}
{\bf N}&=&
 \int d\phi
\frac{\varepsilon_0 {\cal R}}{2 k}\re
\partial_\phi\psi\partial_Z\psi^*{\bf e}_z
\nonumber \\ &=&
\frac{\pi \varepsilon_0 {\cal R}}{k}{\bf e}_z
\im
\sum_{m=-\infty}^\infty
m
\nonumber \\ &&
\left(a^{(1)*}_m H^{(2)}_m +
 a^{(2)*}_m H^{(1)}_m \right)
\nonumber \\ &&
\left(a^{(2)}_m H^{(2)\prime}_m +
 a^{(1)}_m H^{(1)\prime}_m \right)
.
\end{eqnarray}
The final result  (\ref{eq:torque}) follows from
$H^{(1)*}=H^{(2)}$ and
\begin{equation}
\im H_m^{(2)}(Z)H_m^{(1)\prime}(Z)=\frac{2}{\pi Z}.
\end{equation}


\begin{thebibliography}{11}
\bibitem{remark0}
The values are slightly shifted, since there is no amplification outside
the medium.
\bibitem{remark}
Note that $\phi_0$ is the direction to the light source,
i.e., the incoming field actually propagates in direction $\phi_0+\pi$.
With this definition, $\phi_0$ and $\phi$ indeed denote directions
related by time-reversal symmetry.
\end{thebibliography}
\end{document}